\documentclass[twocolumn,prl,preprintnumbers,tightenlines,nofootinbib,superscriptaddress]{revtex4-2}

\usepackage[colorlinks=true,breaklinks=true]{hyperref}
\hypersetup{allcolors=[rgb]{0.0 0.0 1.0},linkcolor=[rgb]{0.75 0.05 0.05}}
\usepackage{orcidlink}
\usepackage{graphicx,wrapfig,float,slashed,subcaption,bbold,bm}
\usepackage{amsmath,amssymb,epsfig,graphicx,xcolor}
\usepackage{epstopdf}
\usepackage{booktabs}
\usepackage{ragged2e}
\usepackage{mciteplus}
\usepackage{mathtools}
\usepackage{enumitem}
\usepackage{cleveref}
\usepackage[normalem]{ulem} % for \sout strikethrough in revision markup

\usepackage{cancel}

\usepackage{makecell}
\usepackage{multirow}
\usepackage{adjustbox}

 \usepackage{comment}

\newcommand{\nc}{\newcommand}
\nc{\non}{\nonumber}
\nc{\hc}{\hbox {h.c.}}
\nc{\noi}{\noindent}
\nc{\barx}{\bar{x}}
\nc{\pbarn}{\;\hbox {pb}}
\nc{\fbarn}{\;\hbox {fb}}

\nc{\hsp}{\hspace{0.5cm}}
\nc{\lsp}{\hspace{1cm}}
\nc{\Lsp}{\hspace{2cm}}
\nc{\LLsp}{\lsp\lsp}
\nc{\lra}{\longrightarrow}
\nc{\p}{\prime}
\nc{\sgn}{\text{sgn}}
\nc{\ph}{\varphi}
\nc{\op}{{\cal O}}

\nc{\beq}{\begin{equation}}  \nc{\eeq}{\end{equation}}
\nc{\bea}{\begin{eqnarray}}  \nc{\eea}{\end{eqnarray}}
\nc{\baa}{\begin{array}}     \nc{\eaa}{\end{array}}
\nc{\bit}{\begin{itemize}}   \nc{\eit}{\end{itemize}}
\nc{\ben}{\begin{enumerate}} \nc{\een}{\end{enumerate}}
\nc{\bce}{\begin{center}}    \nc{\ece}{\end{center}}
\nc{\bpm}{\begin{pmatrix}}   \nc{\epm}{\end{pmatrix}}
\nc{\bvt}{\begin{verbatim}}  \nc{\evt}{\end{verbatim}}
\nc{\bealg}{\begin{equation}\begin{aligned}} 
\nc{\eealg}{\end{aligned}\end{equation}} 

\definecolor{darkgreen}{rgb}{0,0.5,0}

\def\lsim{\mathrel{\raise.3ex\hbox{$<$\kern-.75em\lower1ex\hbox{$\sim$}}}}
\def\gsim{\mathrel{\raise.3ex\hbox{$>$\kern-.75em\lower1ex\hbox{$\sim$}}}}

\def\udots{\mathinner{\mkern1mu\raise1pt\vbox{\kern7pt\hbox{.}}\mkern2mu\raise4pt\hbox{.}\mkern2mu\raise7pt\hbox{.}\mkern1mu}}

\def\dd{\mathrm d}

\newcommand{\tr}{\mathrm{tr}}
\newcommand{\diag}{\mathrm{diag}}
\newcommand{\UV}{\mathrm{UV}}
\newcommand{\IR}{\mathrm{IR}}

\newcommand{\QCD}{\mathrm{QCD}}

\global\long\def\b#1{\left(#1\right)}%
\global\long\def\s#1{\left[#1\right]}%
\global\long\def\ha{\frac{1}{2}}
\global\long\def\pd{\partial}

\begin{document}

\title{$\theta$ Angle and Axial Anomaly in Holographic QCD}

\author{Csaba Cs\'aki\orcidlink{0000-0001-8899-6073}}
\email{csaki@cornell.edu}
\affiliation{Laboratory for Elementary Particle Physics, Cornell University, Ithaca, NY 14853, USA}
\author{Eric Kuflik\orcidlink{0000-0003-0455-0467}}
\email{eric.kuflik@mail.huji.ac.il}
 \affiliation{Racah Institute of Physics, Hebrew University of Jerusalem, Jerusalem 91904, Israel}
\author{Wei Xue\orcidlink{0000-0003-1568-4946}} 
\email{weixue@ufl.edu}
\affiliation{Institute for Fundamental Theory, 
University of Florida, Gainesville, FL 32611, USA}
\author{Taewook Youn\orcidlink{0000-0003-2229-5025}}
\email{taewook.youn@cornell.edu}
\affiliation{Laboratory for Elementary Particle Physics, Cornell University, Ithaca, NY 14853, USA}
\affiliation{School of Physics, Korea Institute for Advanced Study, Seoul 02455, Republic of Korea}

\begin{abstract}
 
We present a bottom-up holographic description of the QCD $\theta$-vacuum and the $U(1)_A$ anomaly in five dimensions. The multi-branched $\theta$-vacuum structure emerges geometrically from a higher-dimensional gauge field, while the axial anomaly is realized through a St\"uckelberg coupling that is dual to a Chern-Simons term. In this framework, the $\eta'$ meson appears as a zero mode of bulk fluctuations, and its mass arises from the anomaly-induced St\"uckelberg term. The construction provides a transparent holographic derivation of the anomaly contribution to the $\eta'$ mass and naturally reproduces the Witten-Veneziano relation between the $\eta'$ mass and the Yang--Mills topological susceptibility.
\end{abstract}

%\date{\today}

\maketitle

\section{Introduction}
Finding a realistic holographic description of QCD has been an important goal ever since the original AdS/CFT correspondence~\cite{Maldacena:1997re}. Soon after Maldacena's original paper, Witten proposed~\cite{Witten:1998uka} a dual description for large-$N$ QCD$_3$ and QCD$_4$, based on type IIB/IIA string theory on particular 10-dimensional backgrounds. Witten also explored the origin of the $\theta$-angle in this setup~\cite{Witten:1998zw}: it arises as a  Wilson loop of a higher-dimensional $U(1)$ gauge field, which naturally explains its periodicity, and also the $\theta$-dependence of the vacuum energy. Sakai and Sugimoto~\cite{Sakai:2004cn} later added quark flavors via 8-branes to Witten's construction, resulting in the Witten-Sakai-Sugimoto (WSS) model. Some effort has also been made to find a low-energy effective description of this model, spearheaded by Bartolini et al. in~\cite{Bartolini:2016dbk}.

In parallel to the top-down approach, % taken by the string theorists 
efforts to construct bottom-up holographic models of QCD were also been initiated~\cite{Erlich:2005qh,DaRold:2005mxj}. Some predictions of those models were very successful, for example reproducing the pion masses and the chiral Lagrangian in the large-$N$ limit, though other features involving high energy behavior seem to require the full-fledged stringy construction~\cite{Csaki:2008dt}. One important issue that should appear in a simple 5D bottom-up model is the vacuum structure and the implementation of the chiral anomalies. Indeed, an early proposal studying the $\eta'$ meson in such a setup was made by Katz and Schwartz~\cite{Katz:2007tf}, where the $\eta'$ state was identified and the validity for  the Witten-Veneziano  relation was argued for. There have been many follow-up papers reusing some of the aspects of the construction of~\cite{Katz:2007tf}, and it remains an active area of research~\cite{Giannuzzi:2021euy,Liu:2026cpr}.   Several previous papers have discussed the  holographic implementation of the axial anomaly using St\"uckelberg-like terms (or their Hodge duals)~\cite{Klebanov:2002gr,Casero:2007ae,Jarvinen:2011qe,Arean:2016hcs,Gursoy:2014ela,Jimenez-Alba:2014iia} from the top-down approach.  Here we present a clean explanation of how this mechanism works in a simple bottom-up 5D holographic model and provide a purely 5D holographic explanation of the Witten-Veneziano relation.

The aim of this paper is to explain the vacuum structure and the chiral anomaly in a the bottom-up 5D holographic QCD model. We will incorporate insights from the WSS model to find a suitable 5D theory that reproduces the expected $\theta$-dependence of the vacuum energy and also incorporates the axial anomalies. While the original string theory based constructions are formulated on a disc, the simple 5D models do not have this ingredient. Nevertheless, some of the topological consequences of the disc are essential, and must be implemented ``by hand" in the 5D model. In particular, the string based model uses the Wilson loop of a bulk gauge field around the disc for the field that couples to the topological ${\rm Tr} F\tilde{F}$ on the boundary; for the 5D construction we simply use the Wilson loop as a full-fledged bulk scalar field $\theta$, with the additional requirement that it be a periodic angular variable. Similarly, the IR boundary condition of this $\theta$ field has to be imposed by hand to match the expectation that the IR brane corresponds to the center of the disc (geometrically the tip of a cone) where all Wilson loops vanish.\footnote{Such an IR boundary condition for the bulk $\theta$ field has previously also been used in~\cite{Gursoy:2007er} arriving from different considerations.} This setup will allow a simple construction of the dual of pure Yang-Mills in the large-$N$ limit. 

Incorporating the anomalies is intuitive as well. The axial $U(1)$ symmetry is represented by a bulk gauge field, which vanishes on the UV brane to ensure it is a global symmetry. The bulk $\theta$ field shifts under this anomalous axial $U(1)$,  resulting in a St\"uckelberg term, and restoring bulk gauge invariance. Chiral symmetry breaking is implemented as usual by a bulk scalar $X$ peaked toward the IR brane. In the absence of the anomalies one finds an exact bulk zero mode, corresponding to the would-be ninth Goldstone boson, the $\eta'$. Turning back on the anomaly generates a mass for the $\eta'$. The Witten-Veneziano relation is automatically and exactly satisfied, which is expected in the large-$N$ limit of QCD. 

The paper is organized as follows. First we review the geometric origin of the $\theta$ angle in string theory constructions of QCD. Next we present the simple 5D model for pure Yang-Mills  and calculate the vacuum energy and the topological susceptibility. The construction of the chiral anomaly in holographic models is explained in the following section. Then we introduce the quark condensate and show the emergence of the $\eta'$ state. We calculate the topological susceptibility and the $\eta'$ mass, and show that it satisfies the Witten-Veneziano relation. 
The appendices are devoted to the matching of the 5D parameters to that of large-$N$ QCD, a discussion of the $R_\xi$-gauge, the discussion of the normalization of the $\eta'$ mode and its relation to the decay constant, and finally to a generalization to more flavors. 

 \section{Geometric Origin of the \texorpdfstring{$\theta$}{theta}-Angle}
 
In holographic QCD models obtained from string theory, the four-dimensional $\theta$ parameter can be naturally realized as the Wilson line of a higher-dimensional gauge field along a compact direction. This geometric origin automatically explains the periodicity of $\theta$.

Here we briefly review the $\theta$-vacuum solution arising from the dual supergravity description in AdS space, following \cite{Witten:1998uka}. The five-dimensional effective field theory in a warped background used in this work is motivated by this construction. Following \cite{Witten:1998uka}, large-$N$ Yang--Mills theory is dual to weakly coupled type IIA string theory with $N$ parallel four-branes. The low-energy effective theory on the branes is a pure $U(N)$ gauge theory, while the branes source a supergravity background with topology
\begin{equation}
\mathcal M = \mathbb{R}^4 \times \text{Disk} \times S^4 \, .
\end{equation}
The geometry is described by the 10D metric \cite{Witten:1998uka}
\begin{align}
\dd s^2 &=
\frac{8\pi}{3}\eta \lambda^3 \sum_{i=1}^{4} (\dd x^i)^2
+ \frac{8}{27}\eta \lambda \pi
\left(\lambda^2 - \frac{1}{\lambda^4}\right) \dd \psi^2
\nonumber \\
&\quad + \frac{8\pi}{3}\eta \lambda
\frac{\dd \lambda^2}{\lambda^2 - \frac{1}{\lambda^4}}
+ \frac{2\pi}{3}\eta \lambda \dd\Omega_4^2 \, .
\end{align}
Here $(\lambda, \psi)$ are polar coordinates on the disk, with $1 \le \lambda < \infty$ and $0 \le \psi \le 2\pi$. 
The parameter $\eta \gg 1$ ensures the validity of the supergravity approximation.

\begin{figure}[t!]
  \centering
  \includegraphics[width=0.65\columnwidth]{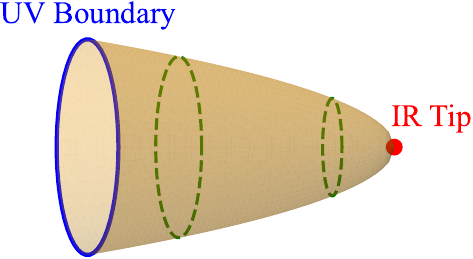}
  \hspace{0.02\columnwidth}
  \caption{\protect\RaggedRight Cigar geometry of the ($\lambda$, $\psi$) directions. The four-dimensional $\theta$-angle is identified with the Wilson loop of $C^{(1)}$ around the circle at the UV boundary. Since the circle  shrinks smoothly to zero size at the tip (IR) of the cigar, the corresponding Wilson loop must be vanishing in the IR. This motivates the IR boundary condition $\theta_\IR=0$ in the effective 5D description.     }
  \label{fig:brane}
\end{figure}

To realize the $\theta$ vacuum~\cite{Witten:1998zw}, we introduce the type-IIA Ramond–Ramond $U(1)$ gauge field $C^{(1)}$ with field strength $W = \dd C^{(1)}$. 
The zero mode of $W$ on the disk, equivalently the Wilson line of $C^{(1)}$ along the $S^1$ UV boundary 
(shown in Fig.~\ref{fig:brane}), gives rise to the $\theta$ angle,  
\begin{equation}
\int_D W = \int_{S^1} C^{(1)} = \theta + 2\pi n \, .
\label{eq:W2_theta}
\end{equation}
The $2\pi n$ ambiguity arises because the left-hand side is a well-defined real number, whereas $\theta$ is an angular variable.
Further, the Wilson line couples to the Yang-Mills fields through Chern-Simons terms on the four-branes,
ensuring that only $\theta \bmod 2\pi$ is gauge invariant. A sketch of the shape of the extra dimensional geometry is shown in Fig.~\ref{fig:brane}. We can see that the topology of the $\lambda ,\psi$ directions is that of a disc, presented here in the shape of a cigar. This type of geometry is typical of gravity duals of confining gauge theories: it also shows up in the Klebanov-Strassler solution~\cite{Klebanov:2000nc} where the conifold singularity of~\cite{Klebanov:2000hb} is smoothed out into a cigar shaped deformed conifold~\cite{Klebanov:2000nc} via the non-perturbative IR effects. In these setups the tip of the cigar represents the location of the IR - the region where confinement happens. In our simplified 5D models we will be representing this via an IR brane. In these models the Wilson loop around the $\psi$ direction will be included as a separate bulk field $\theta$. Clearly as we are reaching the tip of the cigar (the center of the disc), the Wilson loop shrinks to zero, hence the BC capturing the higher dimensional geometry in the 5D model will be $\theta|_{\IR}=0$.

The zero-mode solution of $W$, satisfying the Maxwell equations and boundary conditions, is given by~\cite{Witten:1998zw}
\begin{equation}
W_{\lambda \psi} = (\theta + 2\pi n)\, \frac{3}{\pi \lambda^7} \, .
\end{equation}
The energy associated with this mode,
\begin{equation}
\int d^{10}x \sqrt{|g|}\, W_{\lambda \psi} W^{\lambda \psi} \, ,
\end{equation}
leads to the $\theta$-dependent vacuum energy
\begin{equation}
{\cal E} (\theta) = C \min_{n} (\theta + 2\pi n)^2 \, ,
\end{equation}
where $C$ is a constant independent of $N$.
This quadratic behavior is a well-known feature of large-$N$ QCD, often referred to as the multi-branched vacuum structure.  
Each integer $n$ labels a metastable vacuum branch, and the physical vacuum corresponds to the minimum over all branches. 
This reproduces the expected cusp-like periodic potential found in Witten's analysis of large-$N$ gauge dynamics~\cite{Witten:1979vv,Witten:1998zw}.

\section{Holographic Pure Yang-Mills from 5D and its Vacuum Energy\label{sec:pureYM}}
First we present the simplified model for pure Yang-Mills theories that captures the essential features of the full stringy construction, but is formulated in a 5D warped space with a UV and IR brane, along the lines of the holographic QCD models~\cite{Erlich:2005qh,DaRold:2005mxj}. The background is AdS$_5$ space 
\begin{equation}
\dd s^2 = \frac{R^2}{z^2}\left(\dd x^2 - \dd z^2\right)\, .
\end{equation}
Here $R$ is the AdS curvature radius, and $z_{\UV}<z<z_{\IR}$~\cite{Randall:1999ee}, with $z_{\rm IR}=\Lambda_{\rm QCD}^{-1}\sim {\rm GeV}^{-1}$. Since we are only interested in the topological properties and $\theta_{\QCD}$ dependence, we include only the bulk field $\theta$, which is dual to the topological density ${\rm Tr}\, G\tilde{G}$ of QCD. Based on the previous discussion, this real bulk scalar should really be identified with a Wilson line in a higher dimensional theory, of the sort $\theta \sim \int_{S^1} C^{(1)}$. Hence $\theta$ is an angular variable, with a discrete shift symmetry $\theta \sim \theta +2\pi n$. 

The bulk action for $\theta$ is simply 
\beq
    S_\text{5D} = \int \sqrt{|g|} \frac{g_\theta^2}{2} \partial_M\theta \partial^M \theta = \int \frac{g_\theta^2}{2} \dd\theta \wedge \star \dd\theta.
\eeq
Note that the discrete shift symmetry forbids any non-derivative terms in the Lagrangian. 
The equation of motion is simply that of a free massless scalar in the bulk:
\beq
    \pd_z \s{ \frac{R^3}{z^3} g_\theta^2 \pd_z \theta } = 0.
\eeq
The most important aspect of the theory is the set of boundary conditions (BC's). On the UV boundary the BC is straightforward: $\theta$ is the source for ${\rm Tr}\, G\tilde{G}$,  its UV value must correspond  to $\theta_{\QCD}$ 
\begin{equation}
   \theta|_{\UV} = \theta_{\QCD} 
 + 2 \pi n 
\end{equation}
The $2\pi n$ ambiguity reflects the fact that $\theta|_{\UV}$ is a real-valued field,
whereas $\theta_{\QCD}$ is an angular variable, so their matching is defined only modulo $2\pi$.
On the IR brane the BC is less obvious and requires the full 10D supergravity dual for motivation. As discussed in the previous section, the IR brane corresponds to the tip of the cigar (which is the center of the disc); see FIG.~\ref{fig:brane}. The Wilson loop from which $\theta$ originates must necessarily vanish where the circle smoothly shrinks to zero size. This motivates the IR boundary condition 
\begin{equation}
    \theta|_{\IR} = 0\,.
\end{equation}

 Integrating the equation of motion and applying these boundary conditions yields the background profile:
\beq
    \theta(z) = \frac{\theta_{\QCD}+ 2\pi n }{1 - {z^4_{\UV}}/{z^4_{\IR}}} \b{ 1 - \frac{z^4}{z_{\IR}^4} }.
\eeq
Evaluating the on-shell action on this background profile yields the vacuum energy density $\mathcal{E}$ of pure Yang-Mills:
\beq
    \mathcal{E}_n = \frac{2R^3 g_\theta^2}{z_{\IR}^4} (\theta_{\QCD}+2\pi n)^2\, ,
\eeq
for $z_{\UV} \ll z_\IR$.
The physical vacuum energy is obtained by minimizing over the branches, $n$. 

The topological susceptibility is defined as the second derivative of the vacuum energy density with respect to the $\theta_{\QCD}$ angle. From this holographic profile, we immediately identify the pure Yang-Mills topological susceptibility:
\beq
    \chi_\text{YM} \equiv \frac{\pd^2 \mathcal{E}}{\pd \theta_{\QCD}^2} = \frac{4R^3 g_\theta^2}{z_{\IR}^4}.
    \label{eq:chipureYM}
\eeq
These results establish vacuum energy and structure of pure Yang-Mills, which we will be coming back to in later discussions.

\section{Holographic Axial Anomaly}

One of the most important aspects of the $\theta$ angle in QCD is its relation to the chiral anomaly. Our aim in this section is to explain how this structure is reproduced in the 5D holographic description. 

To incorporate the axial $U(1)_A$ of QCD we  include a 5D bulk gauge field $A$, dual to the axial current $J_\mu^5$. Since this is a global symmetry, we impose Dirichlet BCs for $A_\mu$ on the UV brane, while on the IR brane we impose Neumann BCs. 
Conversely, $A_z$ must satisfy a Dirichlet BC at the IR. However, its behavior at the UV is more nuanced due to the gauge fixing, which will be discussed in a later section.
Later we will introduce the quark condensate responsible for the spontaneous breaking of  $U(1)_A$, but for now we focus only on the anomaly. 

While eventually our theory will be formulated in terms of the bulk scalar $\theta$ (since this is the variable that with a simple UV BC), from the point of view of understanding the structure of the anomalies it is useful to first consider its Hodge dual. In 5D, $\theta$ is dual to a 4-form field strength $H$, which can be written in terms of a 3-form field $C^{(3)}$,
\begin{equation}
    H=\star \dd\theta\,, \qquad  H= \dd C^{(3)}\, .
\end{equation}
The bulk action in terms of the 3-form is $S_\text{dual} = \ha \int \frac{1}{g_\theta^2} H \wedge *H$. 
In this formulation we add the topological Chern-Simons term
\begin{equation}
S_{\rm anomaly}= \int \kappa \, H \wedge A \,,
\end{equation}
capturing the axial anomaly. 

Since our eventual theory should be in terms of $\theta$ rather than $C_3$, we would like to dualize the theory again, but now in the presence of the bulk Chern-Simons term. This can be done in the standard way, 
by imposing the Bianchi identity $dH=0$ via a Lagrange multiplier $\theta$, which will be eventually identified with the bulk scalar:
\beq
    S_H = \int \b{ \frac{1}{2g_\theta^2} H \wedge \star H + \kappa \, A \wedge H + \theta \, \dd H }.
    \label{eq:CS}
\eeq
Integrating the last term by parts, one sees that $H$ couples to the combination $\kappa \, A-\dd\theta$.
In order to maintain gauge invariance, if under a gauge transformation $A \to A + \dd\epsilon$, then $\theta$ must shift by $\theta \to \theta + \kappa \, \epsilon$, just as expected for the $\theta$-parameter. 

We can now go back to the description in terms of the $\theta$ field by integrating out $H$ from (\ref{eq:CS}).
Its equation of motion gives $H = -g_\theta^2 \star(\dd\theta - \kappa \, A)$. Substituting back gives
\beq
    S_\text{5D} = \frac{g_\theta^2}{2} \int \sqrt{|g|}(\partial_M\theta - \kappa \, A_M)  (\partial^M\theta - \kappa \, A^M).
\eeq
The  result is simply a St\"uckelberg Lagrangian, which is manifestly gauge invariant once the shift of the $\theta$ angle is included. The coefficient $\kappa$ if fixed by matching to the QCD axial anomaly, and will be identified with $\kappa= q N_f$, where $q$ is the $U(1)_A$ charge of the quarks, and $N_f$ is the number of flavors.  

Finally, we can see directly how the axial anomaly is reproduced. A global $U(1)_A$ rotation in the 4D CFT theory corresponds  to a bulk $U(1)_A$ gauge transformation whose value on the UV boundary is a constant, $ 
\epsilon(z_{\UV},x)=\alpha $. 
Since this is a gauge transformation, the bulk action is invariant. However, $\theta$ also shifts under the gauge symmetry,
so its UV boundary value, which is indentified with the QCD $\theta$-angle, changes as
\begin{equation}
\theta_{\rm QCD}\to \theta_{\rm QCD}+\kappa \, \alpha.
\end{equation}
This is precisely the expected anomalous shift under a global axial rotation. In this way, the 5D gauge symmetry reproduces the $U(1)_A$ anomaly.

Note that the $C^{(3)}$ 3-form can be thought of as the magnetic dual to $\theta$, hence these two variables are mutually non-local. This implies that if the couplings of $\theta$ are simple local expressions (as we assume it is sourcing ${\rm Tr}G\tilde{G}$ on the UV brane), the couplings of $C^{(3)}$ to the gluon field will be non-local, and one can not very easily identify the effect of the anomaly in the language of the $C^{(3)}$.  

\section{5D Action with Quark Condensate}
To describe spontaneous chiral symmetry breaking, we introduce a complex bulk scalar field $X = \frac{1}{\sqrt{2}} \rho e^{i\phi}$, holographically dual to the quark bilinear operator $\bar{q}q$. The background profile of the modulus $\rho$ encodes the chiral condensate, while the phase $\phi$ contains the would-be Goldstone mode that mixes with the topological angle $\theta$ and with $A_z$. Combining the scalar sector with the gauge-anomaly sector, the full 5D action is\footnote{An alternative implementation of $\chi$SB was presented in~\cite{Arean:2016hcs}.}
\begin{widetext}
\beq
    S = \int \sqrt{|g|} \s{ \frac{g_\theta^2}{2} (\partial_M\theta-\kappa \, A_M)^2  + \ha \rho^2 (\partial_M\phi-q \, A_M)^2  + \ha (\partial_M\rho)^2  - \ha m_X^2 \rho^2 - \frac{1}{4e^2} F^2 + {\rm g.f.}} \,,
\eeq
\end{widetext}

To capture explicit chiral symmetry breaking via the quark mass $m_q$, we introduce a localized potential $S_{\UV}$ on the UV boundary.
\bealg
    S_{\UV} = -\int \dd^4x \sqrt{|g^\text{ind}|} \, V(X), \\
    V(X) \supset - m_q R^{-\frac{3}{2}} X^\dagger + \hc \,, \label{UVterm}
\eealg
where $\sqrt{|g^\text{ind}|} = R^4/z^4$ is the induced 4D metric. 
Since the quark condensate has dimension $\Delta = 3$ we choose for the mass of the bulk scalar $m_X^2 R^2 = -3$.  The profile depends on a spontaneous symmetry-breaking vacuum expectation value $v$ and an explicit symmetry-breaking source $s$. To leading order, it takes the form 
\beq
    \rho(z) = \frac{R^{-3/2}}{\sqrt{2}} \s{ s \, z + v z_{\IR}^{-2} z^3}.
\eeq
We also supplement the action with an $R_{\xi}$ gauge-fixing term to uncouple the longitudinal modes and the scalar phases $\theta$ and $\phi$ from the transverse modes, as detailed in Appendix~\ref{app:rxi}. 

To properly define the global symmetries and the vacuum structure, we must impose boundary conditions consistent with the spontaneous breaking of chiral symmetry by the bulk scalar profile. The $U(1)_A$ symmetry is a global symmetry of the dual field theory, so we require the transverse gauge modes to vanish at the UV boundary. However, the symmetry is broken dynamically in the bulk by the $X$ field, so the gauge field should smoothly satisfy a Neumann condition at the IR wall. The resulting boundary conditions for the gauge fields are
\bealg
    A_\mu|_{\UV} = 0, \\
    \left[ -\frac{1}{e^2} \pd_z \b{ \frac{R}{z} A_z } + \frac{R^3}{z^3} (\rho^2 q \, \phi + g_\theta^2 \kappa \, \theta) \right]_{\UV} = 0, \\
    \pd_z A_\mu|_{\IR} = 0, \quad A_z|_{\IR} = 0. \label{ABC}
\eealg
For the angular variables $\theta,\phi$, the topological angle remains fixed as a non-dynamical source at the UV, while the conjugate momentum for the chiral phase balances the explicit breaking potential:
\bealg
\theta|_{\UV} = \theta_{\QCD}+2\pi n,\\
\quad \frac{R^3}{z^3}\rho^2(\partial_z \phi - q A_z)|_{\UV} = \frac{R^4}{z^4}\frac{\partial V}{\partial \phi}\bigg|_{\UV}\,.
\eealg
For simplicity we will be working in the $n=0$ branch. At the IR boundary, the physics of the anomaly requires a specific mixed boundary compatible with gauge invariance. We therefore imposes the boundary condition on the gauge-invariant linear combination:
\beq
    (q \, \theta - \kappa \, \phi)|_{\IR} = 0.
\eeq
Furthermore, the variation of the action on the IR boundary requires the corresponding conjugate momenta to satisfy
\beq
\left. \frac{R^3}{z^3} \left[ \kappa \, g_\theta^2 (\partial_z \theta - \kappa \, A_z) - q \, \rho^2 (\partial_z \phi - q \, A_z) \right] \right|_{\IR} = 0\,.
\eeq

\section{The $\eta'$ and Topological Susceptibility}

The most immediate consequence of the anomalies and the spontaneous breaking of the axial symmetry is the appearance of the $\eta'$ meson, the would-be $U(1)_A$ goldstone boson %particle that would be the ninth pion 
in the absence of the anomalies.
We are now ready to explore this particle in our holographic setup. The key to understanding this state is to identify the zero mode\footnote{Note, that this zero mode is satisfies the same equations and has the same form as the mode used in warped axion models~\cite{Cox:2019rro}.} one would get if one turned off the anomalies by setting $\kappa\to 0$ (and we are also setting the explicit breaking due to the quark masses to zero $m_q\to 0$). 
In this limit, we should have an exact massless Goldstone boson due to the spontaneously broken axial symmetry, which we can easily identify explicitly. 

The 4D mode zero mode will live partly in $A_z$ and partly in $\phi$, hence it can then be written as 
\beq
    A_z (x,z) = A^{\eta'}_z(z) \eta'(x), \quad \phi(x,z) = \phi^{\eta'}(z) \eta'(x).
\eeq
where $\eta'(x)$ is the four dimensional zero mode satisfying $\Box_4 \eta' (x)=0$.
The bulk equations of motion use the gauge-invariant combinations $\Theta = \pd_z \theta - \kappa \, A_z$ and $\Phi = \pd_z \phi - q \, A_z$. For the physical zero mode, the chiral dynamics require $\Phi = 0$, giving us the simple relation $\pd_z \phi = q \, A_z$. Also, a physical state must make the gauge-fixing constraint vanish, which implies (in the $\kappa\to 0$ limit):
\beq
    -\frac{1}{e^2} \pd_z \b{ \frac{R}{z} A_z } + \frac{R^3}{z^3} \rho^2 q \, \phi = 0. \label{Azconstraint}
\eeq

In the absence of $m_q$ we can approximate the scalar profile by $\rho(z) \simeq \rho_0 z^\Delta$, keeping $\Delta$ general for now, though later we set $\Delta$ to its physical value $\Delta = 3$.
 Plugging this, and the zero-mode constraint, into the Eq.~\eqref{Azconstraint} and applying the IR boundary condition~\eqref{ABC} gives the zero-mode profile:
 \beq
    A^{\eta'}_z(z) = \mathcal{N} z^\Delta \s{ K_\nu(w_{\IR}) I_\nu(w(z)) - I_\nu(w_{\IR}) K_\nu(w(z)) },
\eeq
where $\nu = \frac{\Delta - 1}{\Delta}$, $w(z) = \frac{\beta}{\Delta} z^\Delta$,
and $\beta^2 = q^2 e^2 R^2 \rho_0^2$.
The scalar phase profile directly follows from the $\partial_z\phi= q \, A_z$ constraint:
\beq
    \phi^{\eta'}(z) = \mathcal{N} \frac{q}{\beta} z \s{ K_\nu(w_{\IR}) I_{\nu-1}(w(z)) + I_\nu(w_{\IR}) K_{\nu-1}(w(z)) }.
\eeq

The normalization constant $\mathcal{N}$ is determined by requiring that the 4D kinetic term for this mode is properly normalized, and in terms of the decay constant $f_{\eta'}$ is given by (see Appendix.~\ref{app:feta})
\beq
    \mathcal{N} = -\frac{e^2 \sqrt{2N_f} f_{\eta'}}{R I_\nu(w_{\IR}) \frac{\Gamma(\nu)}{2} \b{\frac{2\Delta}{\beta}}^\nu}.
\eeq

Once we turn the anomaly parameter $\kappa$ back on, it will give a mass to this zero mode.  The anomaly creates a mixing term $(\pd_z \theta - \kappa A_z)^2$ in the bulk action, which will also involve the zero mode $\eta'(x)$. However it will also include a mixing with the entire tower of the $\theta$ modes (the tower of pseudo-scalar spin-0 glueballs), hence one needs to be evaluating the resulting mass term carefully. There is however a simple way to identify the resulting mass, which will also have a very intuitive interpretation in terms of the Witten-Veneziano relation. 

Using the relation for the zero mode $\pd_z \phi = q \, A_z$, we can write the bulk mixing term in the form 
\beq
    (\pd_z \theta - \kappa /q\, \pd_z \phi)^2.
\eeq
The key is that one can completely remove this mixing by a field redefinition for the bulk $\theta$: $\theta(x,z) \to \theta(x,z) - \frac{\kappa}{q} \tilde{\phi}(z) \eta'(x)$. With this field redefinition we completely decouple the $\theta$ from the $A_z,\phi$ fields in the bulk. One also has the {\it pure Yang-Mills} boundary condition on the IR brane for the redefined $\theta$ field: $\theta|_{IR}=0$. Hence all the effects of the anomaly on the $\eta'$ mode 
are pushed into the UV boundary condition:
\beq
    \theta|_{\UV} \to \theta_{\QCD} - \frac{\kappa}{q} \phi^{\eta'}|_{\UV} \eta'(x).
\eeq

Using the boundary value of the wave function $\phi^{\eta'}|_{\UV}$ and the QCD anomaly matching relation $\kappa / q = N_f$ (detailed in the Appendix), the shifted UV $\theta$-angle becomes 
\beq
    \theta|_{\UV} \to \theta_{\QCD} - \frac{\sqrt{2N_f}}{f_{\eta'}} \eta'.
    \label{eq:shiftedtheta}
\eeq
Hence the effect of the anomaly is to produce a mass term for the $\eta'$ mode which is just the $\theta$-dependent part of the pure Yang-Mills energy density with the $\theta$ angle shifted according to (\ref{eq:shiftedtheta}):
\beq
    V_{\rm anomaly}(\eta') = \ha \chi_{\text{YM}} \b{ \theta_{\QCD} - \frac{\sqrt{2N_f}}{f_{\eta'}} \eta' }^2  \,.
\eeq
This will generate the dominant mass of the $\eta'$, and also a VEV that would cancel the entire $\theta$ dependence of the vacuum energy if there were no additional explicit $U(1)_A$ breaking terms. This is of course expected, since we know that for massless fermions the $\theta$ angle is unphysical, and we see explicitly how it would be canceled by a VEV for the $\eta'$ mode. 

Next, we add the explicit chiral symmetry breaking from the UV potential $V(X)$, which is sourced by the quark mass $m_q$. This gives us and additional 4D effective potential for the $\eta'$ field\footnote{Compared to the one-flavor potential in Eq.~\eqref{UVterm}, the overall coefficient is multiplied by $N_f$, reflecting the sum over $N_f$ degenerate flavors.}:
\beq
    V(\eta')_{m_q} \simeq   - N_f m_q v z_{\IR}^{-2} \cos\b{ \frac{2}{\sqrt{2N_f}f_{\eta'}} \eta' - N_f \theta_q} \,,
\eeq
where we have assumed that the dimension of the quark condensate is $\Delta = 3$, and $\theta_q$ is the phase of the quark mass. This $\theta_q$ can be reabsorbed by shifting the $\eta'$ field, which will then make it reappear in the anomaly term, via the invariant CP-violating phase:
\beq
    \bar{\theta} = \theta_{\QCD} - \frac{\kappa}{q} \theta_q \,.
\eeq
The full potential is then written as 
\begin{equation}
V(\eta') \simeq \ha \chi_{\text{YM}} \b{ \bar{\theta} - \sqrt{2N_f} \frac{\eta'}{f_{\eta'}}  }^2 - N_f^2 \chi_q \cos\b{ \sqrt{\frac{2}{N_f}}\frac{\eta'}{f_{\eta'}} }
\label{eq:fullpot}
\end{equation}
with $\chi_q \simeq m_q v /( N_f z_{\IR}^2)$. 

Looking at the quadratic $\eta'$ terms in this potential, we extract the physical meson mass:
\beq
    m_{\eta'}^2 = \frac{2N_f}{f_{\eta'}^2} \b{\chi_{\text{YM}} + \chi_q}.
\eeq
This nicely reproduces the Witten-Veneziano relation~\cite{Witten:1979vv,Veneziano:1979ec}. It shows that the $\eta'$ mass receives a large topological contribution from the pure Yang-Mills susceptibility $\chi_{\text{YM}}$ (the anomaly part), plus the standard explicit mass contribution from the light quarks.

The scaling of these quantities at large $N_c$ and $N_f$, including the resulting $\eta'$ mass, is somewhat nontrivial in the anomaly sector; we discuss it at the end of Appendix~\ref{app:matching}.

Finally, to get the full QCD topological susceptibility $\chi_{\QCD}$, we integrate out the heavy $\eta'$ meson by minimizing $V(\eta')$. This yields
\beq
    \chi_{\QCD} = \b{ \frac{1}{\chi_{\text{YM}}} + \frac{1}{\chi_q} }^{-1}.
    \label{eq:chiQCD}
\eeq
In the exact chiral limit where $m_q \to 0$, the explicit part $\chi_q$ vanishes. This makes the total susceptibility vanish exactly, reproducing the expected result that massless quarks will completely screen the topological vacuum angle.

\section{Topological Susceptibility from Background Analysis}

One can also obtain the vacuum energy and topological susceptibility by examining the adjustment of the background field configurations due to the quark masses. For this we first define the conserved conjugate momenta for the variables $\Theta = \partial_z \theta -\kappa A_z$, $\Phi = \partial_z \phi - q A_z$:
\begin{equation}
\Pi_\theta = \frac{R^3}{z^3} g_\theta^2 \Theta, \qquad \Pi_\phi = \frac{R^3}{z^3} \rho^2 \Phi \,.
\end{equation}
Since the action only depends on the derivatives of the angular variables, the $\Pi_{\theta,\phi}$ are constant in the bulk. We evaluate the total on-shell energy functional in order to find the susceptibility.  The total static potential energy consists not only of the boundary explicit breaking potentials and the angular bulk kinetic energy, but also the radial bulk kinetic energy of the chiral condensate:
\bealg
& E_{\text{total}} = V_{\UV} + \frac{1}{2} \int_{z_{\UV}}^{z_{\IR}} \dd z \left( \frac{z^3}{R^3 g_\theta^2} \Pi_\theta^2 + \frac{z^3}{R^3 \rho^2} \Pi_\phi^2 \right)  \\ 
& \hskip 9em + \int_{z_{\UV}}^{z_{\IR}} \dd z \left[ \frac{R^3}{z^3} (\partial_z \rho)^2 + \frac{R^5}{z^5} m_X^2 \rho^2 \right] \, .
\eealg
Using the bulk equations of motion and integration by parts we find a simple expression for the bulk portion of the energy:
\begin{equation}
 E_{\text{total}} = V_{\UV} +\frac{1}{2} \Pi_\theta^2 \int_{z_{\UV}}^{z_{\IR}} \frac{z^3}{R^3 g_\theta^2} \dd z + \left[ \frac{R^3}{z^3} \rho \partial_z \rho \right]_{z_{\UV}}^{z_{\IR}} \,.
\end{equation}
The integral multiplying the $\Pi_\theta^2$ term is  exactly the inverse of the pure Yang-Mills topological susceptibility: $\chi_{\rm YM}^{-1}$. The remaining radial boundary terms combine with the explicit potential $V_{\UV}$ to reproduce the exact non-linear explicit symmetry breaking potential, $E_{\text{explicit}} = -m_q v z_{\IR}^{-2} \cos(\phi|_{\UV} - \theta_q)$. The total energy functional thus simplifies to
\begin{equation}
E_{\text{total}} \simeq \frac{1}{2\chi_{\rm YM}} \Pi_\theta^2 - m_q v z_{\IR}^{-2} \cos(\phi|_{\UV} - \theta_q) \,.
\label{eq:Etot}
\end{equation}

We would like to write this expression in terms of the physical quantity $\bar\theta$, and find an expression analogous to (\ref{eq:fullpot}). First we can define the integrated phases $\hat{\theta}$ and $\hat{\phi}$ as
\begin{equation}
\hat{\theta} \equiv -\int_{z_{\UV}}^{z_{\IR}} \Theta \, \dd z =\theta_{\UV}-\theta_{\IR}+\kappa \int_{z_{\UV}}^{z_{\IR}}A_z dz= -\frac{\Pi_\theta}{\chi_{\rm YM}} \,,
\end{equation}
\begin{equation}
\hat{\phi} \equiv -(\phi|_{\UV} - \theta_q) - \int_{z_{\UV}}^{z_{\IR}} \Phi \, \dd z = \theta_q-\phi_{\IR} +q \, \int_{z_{\UV}}^{z_{\IR}}A_z dz.
\end{equation}

To eliminate the Wilson line from the above defined $\hat\theta, \hat\phi$ we consider their linear combination:
\begin{equation}
\hat{\theta} - \frac{\kappa}{q} \hat{\phi} = \theta_{\QCD} - \frac{\kappa}{q} \theta_q - \left( \theta|_{\IR} - \frac{\kappa}{q} \phi|_{\IR} \right) .
\end{equation}
The term in the last bracket vanishes due to the IR BCs, hence this combination produces exactly the physically observable $\bar\theta$
\begin{equation}
\hat{\theta} - \frac{\kappa}{q} \hat{\phi} = \theta_{\QCD} - \frac{\kappa}{q} \theta_q = \bar{\theta} \,.
\end{equation}
Finally, in the exact chiral limit $m_q=0$, the Goldstone zero mode satisfies $\Phi=0$. For small $m_q$, we can assume that the background is along this direction, so that corrections for  $\Phi\neq 0$ will be higher order in $m_q$. Therefore, the effective energy potential follows from Eq.~\eqref{eq:Etot},
\begin{equation}
V_{\text{eff}}(\hat{\phi}) \simeq \frac{1}{2} \chi_{\rm YM} \left( \bar{\theta} + \frac{\kappa}{q} \hat{\phi} \right)^2 - m_q v z_{\IR}^{-2} \cos( \hat{\phi} ) \,.
\end{equation}
Essentially $\hat\phi$ plays the role of the VEV of the $\eta'$ mode that we have identified in the previous section, and we are reproducing the same result as in (\ref{eq:fullpot}).
From this effective potential, one can directly read off the total topological susceptibility $\chi_{\QCD}$ and the mass squared for the $\eta^\prime$ meson, exactly as we have done before. 

\section{Conclusion}
Holographic QCD can be approached from two complementary perspectives: a top-down construction based on 10D string theory, 
and a bottom-up description in a 5D warped spacetime. 
In this work, guided by the geometric insights of the top-down picture, we have developed a bottom-up holographic framework that 
gives a transparent understanding of the QCD vacuum structure, the axial anomaly, and the origin of the $\eta'$ mass.

The $\theta$ angle 
arises geometrically from a 
Wilson loop in the higher-dimensional theory. Motivated by this structure, we describe the corresponding degree of freedom in the 5D theory 
by a bulk scalar $\theta$ field. With the appropriate UV and IR boundary conditions, this setup reproduces the multi-branched vacuum 
structure of large-$N$ Yang-Mills theory.

We then incorporate the axial $U(1)_A$ symmetry through a 5D gauge field $A_M$. In this setup, the chiral anomaly is 
encoded by a St\"uckelberg coupling of the form $(\partial_M \theta - \kappa \,A_M)^2$, which can be understood as the dual 
description of a higher-dimensional Chern-Simons interaction between the gauge field and the 3-form field $C^{(3)}$.

Within this framework, the $\eta'$ meson emerges naturally as the would-be Goldstone mode in the limit where the anomaly is turned off. 
Restoring the anomaly lifts this zero mode and generates the $\eta'$ mass. By an appropriate field redefinition, 
the $\eta'$ dependence can be moved into the UV boundary value of the $\theta$ field, making the relation between the $\eta'$ meson, the $\theta$ vacuum, and the anomaly especially transparent.

Our construction provides a clear holographic understanding of the interplay between the $\theta$ vacuum, the axial anomaly, 
and the $\eta'$ meson in QCD. In particular, it shows how the anomaly is incorporated in a simple and gauge-invariant way in 
the 5D description, while naturally reproducing the Witten-Veneziano relation. 
This framework also provides the foundation for studying the QCD axion in holographic QCD \cite{Csaki:2026qjl}, 
where the axion mixes with $\theta$ and $\eta'$ through the same underlying St\"uckelberg structure.

\section{Acknowledgments}
\begin{acknowledgments}
The authors thank Gregory Gabadadze, Liam McAllister, Lisa Randall, Pierre Sikivie, Raman Sundrum, Matt Strassler, Ofri Telem, Charles Thorn for helpful discussions and feedback. We also thank Rashmish Mishra for sending~\cite{Mishra:2026lvq} to us prior to publication. 
CC and TY are supported in part by the NSF grant PHY-2309456. CC and EK are supported in part by grants No 2022713 and 2024091 from the US-Israel BSF. EK is also supported by grant No 2023711 from the US-Israel BSF. TY is also supported in part by the Samsung Science and Technology Foundation under Project Number SSTF-BA2201-06. WX is supported in part by the U.S. Department of Energy under grant DE-SC0022148 at the University of Florida.
Part of this research was performed at the Munich Institute for Astro-, Particle and BioPhysics (MIAPbP)  funded by the  DFG  grant EXC-2094 – 390783311, and in part at the Aspen Center for Physics, supported by the NSF grant PHY-2210452 and the  Simons Foundation grant 1161654. EK is grateful to Cornell University for its hospitality and support during a sabbatical visit.

\end{acknowledgments}

\section{Note Added}
After finishing this work we learned of~\cite{Mishra:2026lvq} which also investigated the effects of the $\theta$ angle in 5D holographic models of the strong interactions, and used a setup similar to our pure Yang-Mills construction.   

\appendix
\setcounter{secnumdepth}{2}

\section{QCD Matching}
\label{app:matching}
\paragraph{{\bf Bulk Gauge Coupling ($g_\theta^2$)}}
The coupling for the $\theta$ field is determined by the two-point correlation function of the topological charge density in pure Yang-Mills theory~\cite{Katz:2007tf}. The perturbative result at large momentum $Q$ is $\chi_\text{top}(Q^2) \propto -N_c^2 \frac{\alpha_s^2}{32\pi^4} Q^4 \ln Q^2$. Calculating the holographic correlator from the bulk action $\int \frac{g_\theta^2}{2} (\dd \theta)^2$ and matching the coefficients yields
\beq
g_\theta^2 = \frac{N_c^2 \alpha_s^2}{4\pi^4 R^3}.
\eeq
Note however, that this matching is purely perturbative, and applies in the $Q^2\gg 1/z^2_{\IR}$ limit. A large non-perturbative running is expected to correct the UV value of $g_\theta$ to the IR value relevant for the topological susceptibility and the $\eta'$ mass. Hence from the point of view of these quantities $g_\theta$ is essentially a free parameter.

\paragraph{{\bf Anomaly Coefficients ($\kappa$)}}
The divergence of the singlet axial current sums the anomaly contributions from all $N_f$ flavors: $\partial_\mu J^\mu_5 = 2N_f \frac{\alpha_s}{8\pi} G\tilde{G}$. Under a chiral rotation by angle $\alpha$, the path integral measure shifts the vacuum angle by $\theta_{\QCD} \to \theta_{\QCD} + 2N_f \epsilon$. In the holographic dual, a bulk gauge transformation $\kappa_B$ shifts the St\"ueckelberg field by $\delta \theta = \kappa \, \epsilon$. Identifying the transformation properties implies the matching condition:
\beq
\kappa = q \, N_f.
\eeq

\paragraph{{\bf Gauge Couplings ($e^2$)}}
The coupling for the bulk axial gauge field is fixed by the two-point function of the flavor-singlet axial current $J_\mu^5$. The perturbative fermion loop calculation gives $\Pi_A(Q^2) = -\frac{N_c N_f}{12\pi^2} \ln Q^2$. Matching this to the holographic result derived from the bulk gauge kinetic term $-\frac{1}{4e^2} F^2$ determines the coupling:
\beq
e^2 = \frac{6\pi^2 R}{N_c N_f}.
\eeq
This coupling also determines the decay constant for the singlet meson
\beq
\sqrt{2N_f} f_{\eta'} = \lim_{z \to z_{\UV}} \frac{R}{e^2} \frac{A _z}{z}\,.
\label{eq:feta}
\eeq

\paragraph{{\bf $U(1)$ Charges ($q$)}}
The charge $q$ is determined by the transformation of the chiral condensate $\langle \bar{q} q \rangle$ under the axial symmetry. Since $\langle \bar{q} q \rangle \to e^{2i\epsilon} \langle \bar{q} q \rangle$, and the dual bulk scalar $X$ transforms as $X \to e^{iq\epsilon} X$, we identify the charge:
\beq
q = 2.
\eeq

\paragraph{{\bf Large-$N_c$ and Large-$N_f$ Scaling}}
The matchings above also fix the large-$N_c$ and large-$N_f$ scalings of the various couplings and observables, which are somewhat nontrivial in the anomaly sector. The anomaly coefficient is $\kappa = q N_f \sim \mathcal{O}(N_f)$, the axial gauge coupling is $e^2 = 6\pi^2 R/(N_c N_f) \sim \mathcal{O}(1/(N_c N_f))$, and the bulk $\theta$ coupling is $g_\theta^2 = N_c^2 \alpha_s^2/(4\pi^4 R^3) \sim \mathcal{O}(N_c^0)$, where we used the 't~Hooft scaling $\alpha_s \sim 1/N_c$. Consequently, the pure Yang-Mills topological susceptibility $\chi_{\text{YM}} = 4 R^3 g_\theta^2/z_{\IR}^4 \propto g_\theta^2$ scales as $\mathcal{O}(N_c^0)$, as expected for a pure-glue quantity, while the singlet decay constant fixed by~\eqref{eq:feta} scales as $f_{\eta'}^2 \sim \mathcal{O}(N_c)$. Inserting these into the Witten-Veneziano relation, the anomaly contribution to the $\eta'$ mass scales as
\beq
m_{\eta'}^2 \sim \frac{N_f \chi_{\text{YM}}}{f_{\eta'}^2} \sim \frac{N_f}{N_c} \,.
\eeq
Therefore, in the 't~Hooft limit ($N_c \to \infty$ with $N_f$ fixed) the $\eta'$ mass squared scales as $\mathcal{O}(1/N_c)$ and vanishes, whereas in the Veneziano limit ($N_c, N_f \to \infty$ with $x = N_f/N_c$ fixed) it remains finite, $m_{\eta'}^2 \sim \mathcal{O}(1)$.

\section{$R_\xi$ Gauge Fixing}
\label{app:rxi}
To rigorously define the physical zero modes and justify the exact constraints utilized in the main text, we must eliminate the mixing between the longitudinal gauge components and the scalar phases. This is achieved by introducing a generalized $R_\xi$ gauge-fixing term to the bulk action. Focusing on the $U(1)_A$ chiral sector for clarity, the gauge-fixing action is given by
\bealg
S_\text{g.f.} =& -\frac{1}{2e^2 \xi} \int d^5x \sqrt{|g|} \left[ \frac{z^2}{R^2} \partial_\mu A^\mu - \xi \frac{z^3}{R^3} \partial_z \left(\frac{R}{z} A_z\right) \right. \\ 
&  \hskip 9em + e^2 \xi (g_\theta^2 \kappa \, \theta + \rho^2 q \, \phi) \bigg]^2\,.
\eealg
The terms multiplying $\xi$ define the gauge constraint $\mathcal{F}_A$
\beq
\frac{\mathcal{F}_A}{\xi} = -\frac{1}{e^2} \partial_z \left(\frac{R}{z} A_z\right) + \frac{R^3}{z^3} \rho^2 q \, \phi + \frac{R^3}{z^3} g_\theta^2 \kappa \, \theta\,.
\eeq
In $R_\xi$ gauge, the transverse and longitudinal components decouple. 

\section{Kinetic Normalization\label{app:feta}}
To properly normalize the 4D effective fields, we evaluate the exact 5D kinetic normalization integral $N$, which is a functional of the dynamical fluctuations around the vacuum background:
\beq
N = \int_{z_{\UV}}^{z_{\IR}} \dd z \left( \frac{R}{e^2 z} (A_z^{\eta'})^2 + \frac{R^3}{z^3} \rho^2 (\phi^{\eta'})^2 \right)\,.
\eeq
Using the exact gauge-invariant relation for the fluctuations $\partial_z \phi^{\eta'} = q A_z^{\eta'} + \Phi^{\eta'}$ (where $\Phi^{\eta'} = \partial_z \phi^{\eta'} - q A_z^{\eta'}$), along with the vanishing gauge-fixing constraint, we can integrate the scalar term by parts. The $q A_z^{\eta'}$ piece perfectly cancels the gauge kinetic term, leaving
\beq
N = \left[ \frac{1}{q e^2} \frac{R}{z} A_z^{\eta'} \phi^{\eta'} \right]_{z_{\UV}}^{z_{\IR}} - \int_{z_{\UV}}^{z_{\IR}} \dd z \frac{1}{q e^2} \frac{R}{z} A_z^{\eta'} \Phi^{\eta'}\,.
\eeq
The IR boundary evaluation vanishes due to the Dirichlet condition $A_z^{\eta'}|_{\IR} = 0$. At the UV boundary, we apply the holographic flux definition of the decay constant \eqref{eq:feta}. Because the explicit symmetry-breaking fluctuation $\Phi^{\eta'}$ is localized near the UV boundary, within the region where the longitudinal gauge flux $A_z^{\eta'}/z$ remains flat, we can extract the boundary flux limit from the integral operator. This allows us to identify the physical, gauge-invariant 4D effective phase fluctuation $\hat{\phi}^{\eta'}$, defined as the total integrated response across the bulk:
\beq
\hat{\phi}^{\eta'} \equiv - \phi^{\eta'}|_{\UV} - \int_{z_{\UV}}^{z_{\IR}} \dd z \, \Phi^{\eta'}\,.
\eeq
Enforcing canonical normalization ($N = 1$) rigidly locks this physical phase:
\beq
\hat{\phi}^{\eta'} = \frac{q}{\sqrt{2N_f} f_{\eta^\prime}}\,.
\eeq

In the strict chiral limit ($m_q \to 0$), the explicit symmetry-breaking fluctuation vanishes identically ($\Phi^{\eta'} = 0$), meaning the physical phase is entirely determined by the boundary value ($\hat{\phi}^{\eta'} = - \phi^{\eta'}|_{\UV}$). By requiring that the 4D kinetic term of the zero mode is properly normalized ($N=1$), the normalization constant $\mathcal{N}$ is determined entirely by this UV boundary term:
\beq
\mathcal{N} = \left[ - \lim_{z \to z_{\UV}} \frac{R}{q_A e_A^2 z} \left( \frac{A_z^{\eta'}(z)}{\mathcal{N}} \right) \left( \frac{\phi_X^{\eta'}(z)}{\mathcal{N}} \right) \right]^{-1/2}\,.
\eeq
Substituting the explicit zero-mode profiles into this boundary term yields
\beq
\mathcal{N} = \frac{q e^2 \sqrt{2N_f} f_{\eta'}}{R \Big[ K_\nu(w_{\IR}) I_\nu(w_{\UV}) - I_\nu(w_{\IR}) K_\nu(w_{\UV}) \Big]}\,.
\eeq
In the limit $z_{\UV} \to 0$, the $I_\nu(w_{\UV})$ term mathematically vanishes. Substituting the small-argument expansion $K_\nu(w_{\UV}) \simeq \frac{\Gamma(\nu)}{2} \left(\frac{w_{\UV}}{2}\right)^{-\nu}$ reveals a perfect cancellation of the remaining $z_{\UV}$ dependence, yielding the exact constant:
\beq
\mathcal{N} = - \frac{e^2 \sqrt{2N_f} f_{\eta'}}{R \, I_\nu(w_{\IR}) \frac{\Gamma(\nu)}{2} \left(\frac{2\Delta}{\beta}\right)^\nu} \,.
\eeq

\section{Extension to Two Flavors}

To generalize the holographic framework to the realistic case of two light flavors ($N_f=2$), the bulk gauge symmetry must be enlarged to include the  $SU(2)_L\times SU(2)_R$ sector. The scalar field $X$ is promoted to a matrix-valued field transforming as a bifundamental:
\beq
    X = \frac{1}{\sqrt{2}}\rho \, Ue^{i\phi} \,.
\eeq
Here, $U = \exp(i\pi^a T^a)$ encodes the pion triplet, and the modulus $\rho = \diag(\rho_u, \rho_d)$ describes the chiral condensate for both quark flavors.

The topological susceptibility of the theory is determined by the explicit breaking of the chiral symmetries on the UV boundary. We introduce a boundary potential involving the quark mass matrix $M_q = \diag(m_u, m_d)$ with complex phases $\theta_u$ and $\theta_d$:
\beq
    V(X) \supset -\tr\b{ M_q^\dagger X } + \hc \,.
\eeq
Assuming isospin symmetry of the background condensate profile at the boundary ($\rho_u \simeq \rho_d = v z_{\IR}^{-2}$), this trace expands into the sum of two cosine terms representing the independent flavor contributions:
\bealg
    -v z_{\IR}^{-2} & \left[ m_u \cos\b{ \phi|_{\UV} + \pi|_{\UV}^3 - \theta_u } \right. \\
    & \hskip 8em + \left. m_d \cos\b{ \phi|_{\UV} - \pi|_{\UV}^3 - \theta_d } \right] \,.
\eealg
To find the physical vacuum, we minimize this explicit breaking boundary energy with respect to the neutral pion phase $\pi|_{\UV}^3$, yielding the condition:
\beq
    m_u \sin\b{ \phi|_{\UV} + \pi|_{\UV}^3 - \theta_u } = m_d \sin\b{ \phi|_{\UV} - \pi|_{\UV}^3 - \theta_d } \,.
\eeq
Once this dynamical adjustment of the pion field is substituted back into the boundary energy, the effective potential for the flavor-singlet phase $\phi_{\UV}$ will depend on the invariant combination of the background phases $\theta_q = \theta_u + \theta_d = \arg \det M_q$:
\bealg
    V_{\rm eff}(\phi) &= -v z_{\IR}^{-2} \b{ m_u + m_d } \\
    & \hskip 4em \times \sqrt{ 1 - \frac{4 m_u m_d}{(m_u + m_d)^2} \sin^2\b{ \phi|_{\UV} - \frac{\theta_q}{2} } } \,.
\eealg
Expanding this potential for small fluctuations reveals the explicit chiral symmetry breaking part of the topological susceptibility, $\chi_q$:
\beq
    \chi_q = \frac{m_u m_d}{m_u + m_d} v z_{\IR}^{-2}
\eeq
This demonstrates a crucial feature of the two-flavor vacuum: when minimizing the energy, the neutral pion $\pi^3$ dynamically adjusts to absorb the isospin-breaking difference between the up and down quarks. Consequently, the contribution to the flavor-singlet sector is not additive but is dominated by the lightest quark flavor through the reduced mass. This perfectly reproduces the explicit mass contribution expected from chiral perturbation theory without.

Generalizing this framework to a larger number of flavors, $N_f > 2$, is straightforward. It involves expanding the bulk gauge group to $SU(N_f)_L\times SU(N_f)_R$ and incorporating the full set of $N_f^2 - 1$ meson generators, following the identical algebraic procedure to integrate out the massive non-singlet modes and isolate the singlet dynamics. The result will be 
\begin{equation}
    \chi_q= \big(\, \sum_{i=1}^{N_f} \frac{1}{m_i}\, \big)^{-1} v z_{\IR}^{-2}
\end{equation}
Again as expected the lightest quark will dominate the quark contribution, and via (\ref{eq:chiQCD}) also the total susceptibility. 
The contribution of the anomaly sector stays is independent of the number of flavors, as the topological mass generation continues to be governed entirely by the Abelian $U(1)_A$ gauge field.

\bibliography{bibtex}

\onecolumngrid

\end{document}